\documentclass[conference]{IEEEtran}
\usepackage{amsmath, amssymb, bm, cite, epsfig, psfrag}
\usepackage{graphicx}
\usepackage{soul} 
\usepackage[margin=0.56in]{geometry}
\usepackage{dblfloatfix}
\usepackage{array}
\usepackage{lipsum}

\usepackage[font=small]{caption}
\usepackage{epstopdf}	
\usepackage{longtable}
\usepackage{supertabular,booktabs}
\usepackage{bbm}
\usepackage{multirow}
\usepackage[usenames,dvipsnames]{xcolor}

\usepackage{subfigure}
\usepackage{etoolbox}
\usepackage{pbox}
\usepackage{fixltx2e}
\usepackage{tabu}	
\usepackage{filecontents}
\usepackage{enumerate}
\usepackage{textcomp}
\usepackage{colortbl}
\usepackage{fancyhdr}
\usepackage{bm}

\newtoggle{conference}

\togglefalse{conference} 
\graphicspath{{figures/}}

\newcolumntype{?}{!{\vrule width 2pt}}

\newcolumntype{P}[1]{>{\centering\hspace{0pt}}p{#1}}
\newcolumntype{M}[1]{>{\centering\hspace{0pt}}m{#1}}
\newcolumntype{L}{>{\centering\arraybackslash}m{3cm}}

\fancypagestyle{firststyle}{
	\fancyhf{}
	\fancyhead[L]{ \vspace{+0.6 cm} \small Y. Xing and T. S. Rappaport, ``Millimeter Wave and Terahertz Urban Microcell Propagation Measurements and Models (Invited Paper),'' 2021 \textit{IEEE Communications Letters}, Sept. 2021, pp. 1-5.
	}     

}
\IEEEoverridecommandlockouts

\begin{document}
\title{Millimeter Wave and Terahertz Urban Microcell Propagation Measurements and Models}
\author{\IEEEauthorblockN{Yunchou Xing, and Theodore S. Rappaport (Invited Paper)}

\IEEEauthorblockA{	\small NYU WIRELESS, NYU Tandon School of Engineering, Brooklyn, NY, 11201, \{ychou,  tsr\}@nyu.edu\\
	 }\vspace{-1.0 cm}
					\thanks{}
}

\maketitle
\thispagestyle{firststyle}

\begin{abstract} 
Comparisons of outdoor Urban Microcell (UMi) large-scale path loss models, root mean square (RMS) delay spreads (DS), angular spreads (AS), and the number of spatial beams \textcolor{black}{for extensive measurements performed at 28, 38, 73, and 142 GHz are presented in this letter. Measurement campaigns were conducted from} 2011-2020 in downtown Austin, Texas, Manhattan (New York City), and Brooklyn, New York with communication ranges up to 930 m. \textcolor{black}{Key similarities and differences in outdoor wireless channels are observed when comparing the channel statistics across a wide range of frequencies from millimeter-wave to sub-THz bands.} Path loss exponents (PLEs) are remarkably similar over all measured frequencies, when referenced to the first meter free space path loss, and the RMS DS and AS decrease as frequency increases. The similar PLEs from millimeter-wave to THz frequencies imply that \textcolor{black}{spacing between cellular base stations} will not have to change as carrier frequencies increase towards THz, since wider bandwidth channels at sub-THz or THz carrier frequencies will cover similar distances because antenna gains increase quadratically with increasing frequency when the physical antenna area remain constant. 

\end{abstract}
    


\section{Introduction}~\label{sec:intro}


Inspired by the success of 5G commercial deployments at millimeter wave (mmWave) and sub-6 GHz frequencies, futuristic wireless communication systems (e.g., 6G and beyond) will likely utilize sub-THz and THz frequencies above 100 GHz to provide not only much higher data rates (Tbps) with near-zero latency \cite{rappaport19access,xing21a,Elayan20survey}, but also global coverage with unmanned aerial vehicles (UAV), high altitude platform stations (HAPS), satellites, terrestrial and maritime stations \cite{liu20a}, ushering in innovative applications such as autonomous vehicles/drones, wireless cognition, and precise localization with centimeter-level accuracy \cite{liu20a,Kanhere20a}. Fixed point-to-point communications such as wireless fronthaul and backhaul (e.g., integrated access and backhaul, and Xhaul), operating at data rates on the order of Tbps, will enable fiber replacement in rural areas to provide wireless coverage and services to the less populated areas, and enable wireless replacements of edge data centers \cite{rappaport19access,viswanathan20A, ghosh195g,Rap17a}. 

Joint communication and radar sensing \cite{ali20sensing,dokhanchi19a} will be an important feature of 6G wireless communication systems, where both active sensors such as radar/Lidar and passive sensors like cameras and spectroscopy \cite{rappaport19access} will support new applications like automotive radars, gesture and activity recognition, and contextual awareness \cite{ali19automotive, liu20a}. Sensing aided communications will support beam configuration and alignment, as well as challenging mobility applications like autonomous systems, and communication functionality in radar sensing will help with high-resolution imaging and precise localization with centimeter-level accuracy \cite{ali20sensing,Kanhere20a}.


Work in \cite{Ju20a,Xing21b} presented an indoor wideband measurement campaign in a hotspot office environment at 142 GHz with both line-of-sight (LOS) and non-LOS (NLOS) scenarios, looking at large commercially relevant distances up to 40 m. To date, previous outdoor propagation measurements primarily focus on LOS scenarios using either reflected materials \cite{ma18channel} or an RF-over-fiber extension \cite{abbasi20ICC} of a vector network analyzer based sounding system with coverage ranges of 100 m in outdoor urban environments. Recent work in \cite{xing21icc} presented 142 GHz outdoor Urban Microcell (UMi) radio propagation measurements for LOS and NLOS scenarios with coverage ranges up to 117 m.

This letter presents a comprehensive comparison of outdoor urban wireless channels using transmitting base station antenna heights that varied between 4 and 36 m above ground in four frequency bands from 28 to 142 GHz in UMi environments for both LOS and NLOS scenarios with coverage ranges up to 930 m, based on extensive outdoor radio propagation measurements conducted from 2011 to 2020. \textcolor{black}{Compared to our previous work in [17-21], which presented outdoor UMi channel statistics of path loss and RMS delay spread, this letter introduces new findings of outdoor time and angular statistics (RMS AOA/AOD spread, average number of AOA/AOD directions, and RMS delay spread) at frequencies from 28-142 GHz, which will be useful for multiple-input multiple-output (MIMO) rank analysis and capacity predictions for future sub-THz systems.} Sections \ref{sec:140PL}, \ref{sec:DS}, and \ref{sec:AS} compare both outdoor UMi directional and omnidirectional large-scale path loss models, root mean square (RMS) delay spread and angular spread across the aforementioned four frequency bands. Section \ref{conclusion} concludes the similarities and differences of the time and spatial channel characteristics over frequencies from 28 to 142 GHz.







\begin{table}[]\caption{Outdoor UMi measurement campaigns at 28, 38, 73, and 142 GHz \cite{rappaport2013millimeter,Mac17JSACb,xing18GC,Xing21b}. }\label{tab:sounder}
	\centering
	\begin{tabular}{llllll}
		\hline
		\multicolumn{1}{|l|}{\textbf{RF Freq.}}   & \multicolumn{1}{c|}{\textbf{RF}} &  \multicolumn{1}{l|}{\textbf{Antenna}} & \multicolumn{1}{c|}{\textbf{Ant. Gain}} & \multicolumn{1}{c|}{\textbf{Campaign}}\\ 
		
		\multicolumn{1}{|c|}{\textbf{(GHz)}}   & \multicolumn{1}{c|}{ \textbf{Bandwidth}} &  \multicolumn{1}{c|}{\textbf{HPBW}} & \multicolumn{1}{c|}{\textbf{(dBi)}} & \multicolumn{1}{c|}{\textbf{(UMi)}}  \\ \hline
		
		\multicolumn{1}{|c|}{28  \cite{rappaport2013millimeter}}           & \multicolumn{1}{c|}{0.8 GHz }       & \multicolumn{1}{c|}{10.9\textdegree}           & \multicolumn{1}{c|}{24.5}   & \multicolumn{1}{c|}{Manhattan 2012}            \\ \hline
		
		\multicolumn{1}{|c|}{38  \cite{rappaport2015wideband}}           & \multicolumn{1}{c|}{0.8 GHz }       & \multicolumn{1}{c|}{7.8\textdegree}           & \multicolumn{1}{c|}{25.0  }    & \multicolumn{1}{c|}{Austin 2011}           \\ \hline
		
		\multicolumn{1}{|c|}{73  \cite{Mac17JSACb}}      & \multicolumn{1}{c|}{1.0 GHz}     & \multicolumn{1}{c|}{7.0/15.0\textdegree}           & \multicolumn{1}{c|}{27.0/20.0}  & \multicolumn{1}{c|}{Brooklyn 2016}             \\ \hline
		
		\multicolumn{1}{|c|}{142  \cite{xing21icc}}         & \multicolumn{1}{c|}{1.0 GHz}      & \multicolumn{1}{c|}{8.0\textdegree}            & \multicolumn{1}{c|}{27.0}  & \multicolumn{1}{c|}{Brooklyn 2020}             \\ \hline
	\end{tabular}
\vspace{-0.5cm}
\end{table}

\section{UMi Large-Scale Path Loss and  Models in 28, 38,  73, 142 GHz Bands}\label{sec:140PL}

Understanding the wireless channels above 100 GHz is the critical first step for researchers to design future THz communication systems for 6G and beyond. This paper analyzes and compares  outdoor UMi propagation measurements at 28 GHz \cite{rappaport2013millimeter} (Manhattan, New York), 38 GHz (Texas, Austin) \cite{rappaport2013broadband}, 73 GHz (Brooklyn, New York) \cite{Mac19COMP}, and most recently 142 GHz (Brooklyn, New York) \cite{xing21icc}, conducted over a nine-year period since 2011 with communication ranges up to 930 m \cite{rappaport2015wideband,Mac15c,rappaport2013millimeter,Mac19COMP,Sun14a,Sun16b,xing21icc}. \textcolor{black}{Compared to previous measurements at frequencies below 100 GHz in [17-21], more challenges exist when building channel sounders at 142 GHz \cite{xing18GC}. The limited transmit power and severe free space path loss in the first meter at higher frequencies require the channel sounder system to use high gain directional antennas/arrays which need longer scan time to capture signals from all possible directions.} The measurement locations and procedures at each frequency can be found in \cite{rappaport2015wideband,Mac15c,rappaport2013millimeter,Mac19COMP,Sun14a,Sun16b,xing21icc}. Wideband sliding correlation-based (time domain spread spectrum) channel sounding systems with identical rotatable horn antennas for each frequency at both link ends were used, as shown in Table \ref{tab:sounder}.

\textcolor{black}{In prior work \cite{rappaport2013millimeter,rappaport2013broadband} at 28 and 38 GHz, the close-in (CI) free space reference distance path loss model with free space reference distances of $d_0=$ 3 m and 5 m were used. Subsequent work in \cite{rappaport2015wideband,Sun16b} studied the optimized reference distance $d_{opt}$ and showed using a $d_0 =$ 1 m free space reference distance has negligible difference in accuracy when compared to the optimal $d_{opt}$, \textcolor{black}{ and is more sensible for a universal standard for comparing path loss among different frequencies, locations, and researchers \cite{Sun16b}}. In addition, using the 1 m close-in free space reference distance (a leverage point) assures path loss has a continuous physical tie to signal strength over distance for multi-frequency bands and wide ranges of environments \cite{Sun16b}. Therefore, the CI path loss model with a $d_0=$ 1 m reference distance (the equations could be found in \cite{Mac15b,rappaport2015wideband,3GPP2019,Ju20a,Rap17a,Sun16b}) is used to present the measured path loss and shadow fading in this paper. The methods to compute the multipath numbers, the corresponding channel-gain, RMS delay spread, and RMS angular spread can be found in \cite{Ju20a}.}

\subsection{UMi CI Path Loss Model in a Single Frequency Band}

The outdoor UMi LOS directional path loss exponent (PLE) for the single-frequency CI model is $n=$ 2.3 at 28 GHz, $n=$ 1.9 at 38 GHz, $n=$ 2.0 at 73 GHz, and $n=$ 2.1 at 142 GHz with shadow fading standard deviations of $\sigma=$ 4.3 dB, 3.5 dB, 1.9 dB, and 2.8 dB as shown in Table \ref{tab:PLcomp}, respectively. The higher PLE at 28 GHz is because of antenna misalignment or foliage attenuation between the TX and RX, indicating that accurate beam searching and beam steering algorithms are needed for directional antennas. The LOS measurements show that there is only 1-2 dB larger average loss per decade of distance in the 142 GHz band compared to the average path loss in the 38 and 73 GHz bands when referenced to the first meter free space path loss \cite{Sun16b,rappaport19access,Sun14a}, which can be easily compensated for in a practical system by higher antenna gains. 

The higher gain antennas will not take a larger physical area since the antenna gain increases quadratically as frequency increases if the physical size of the antenna (effective aperture) is kept constant over frequency \cite{rappaport19access,Mac17JSACb,xing18GC}. Thus, the infrastructure spacing in mobile systems will not have to change as frequencies increase up to THz scale, since increased bandwidth channels at higher frequencies can serve over similar distances if the antenna area remains constant \cite{rappaport19access,Mac17JSACb,xing18GC}.


$\text{NLOS}_{\text{Best}}$ denotes the optimal situation at each NLOS measurement location that antennas of the TX and RX are pointing in the best direction when the RX captures the maximum power \cite{Xing21b,Mac15b,Ju20a}. This measurement approach emulates how practical systems would employ directional beamforming in a real-world link to maximize the signal-to-noise ratio (SNR). 

The $\text{NLOS}_{\text{Best}}$ PLEs of the CI model are $n=$ 3.8, 2.7, 3.1, and 3.1 at 28, 38, 73, and 142 GHz, respectively, as shown in Table \ref{tab:PLcomp}. \textcolor{black}{The buildings are less dense in Austin (38 GHz) than in downtown Manhattan (28 GHz) and Brooklyn (73 and 142 GHz), thus the transmitted signal at 38 GHz did not encounter as many physical obstructions (blockages) as in Manhattan and Brooklyn, resulting in a substantially lower PLE of 2.7 at 38 GHz compared to the PLEs at the other three frequencies. However, the $\text{NLOS}_{\text{Best}}$ PLEs at 28, 73, and 142 GHz are remarkably similar to one another with slightly lower PLEs at higher frequencies, which is because of the stronger reflections at higher frequencies (which is also observed in indoor measurements at 28, 73, and 142 GHz \cite{xing19GC,Xing21b}).}

\begin{figure}    
	\centering
	\includegraphics[width=0.45\textwidth]{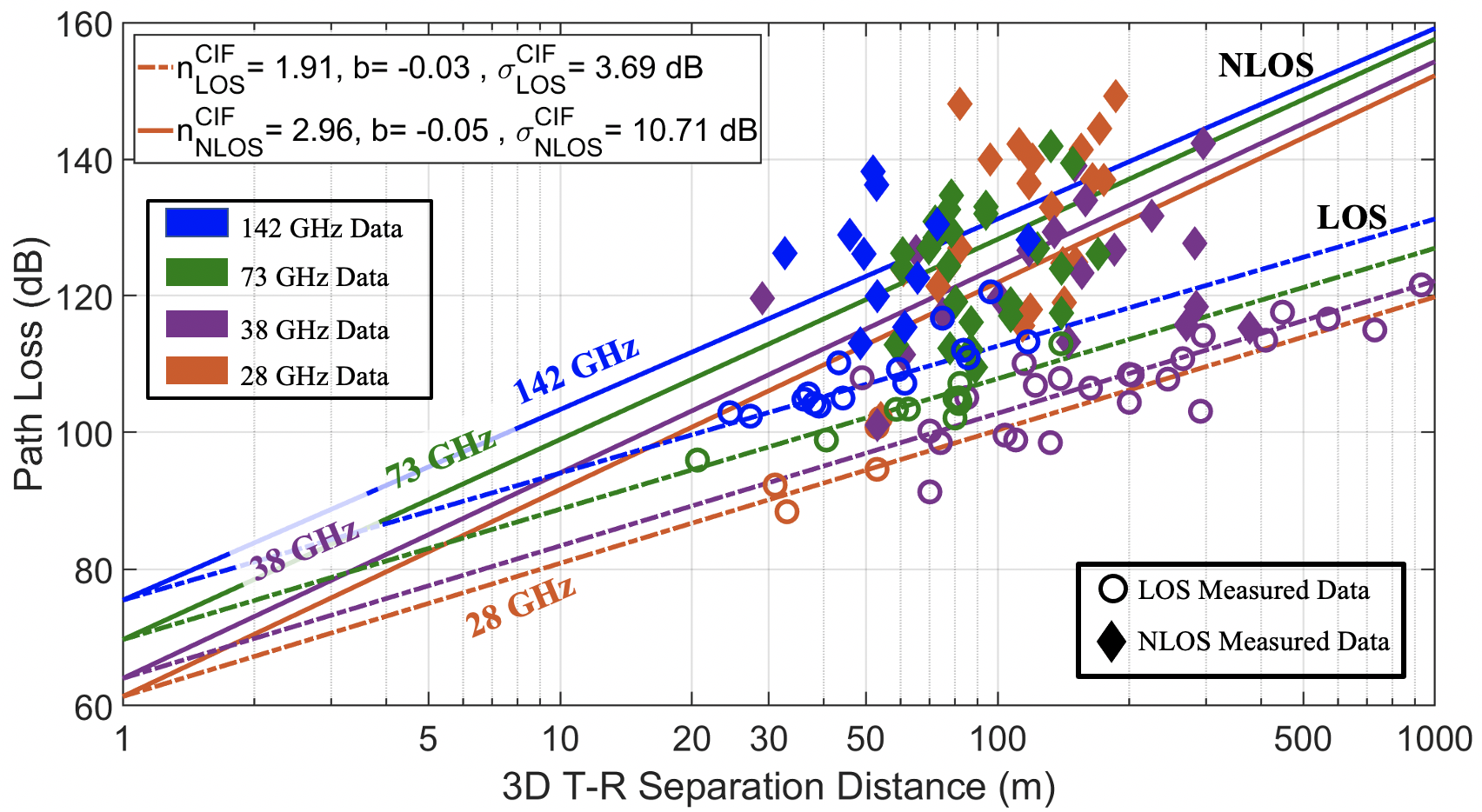}
	\caption{Outdoor UMi 28, 38, 73, and 142 GHz multi-band omnidirectional CIF path loss models with 1 m free space reference distance and without antenna gains \cite{Sun16b,Mac15c,Samimi15b,rappaport2015wideband,Mac19COMP,xing21icc}. The $f_0$ computed by \eqref{equ:CIF} is 73 and 62 GHz for LOS and NLOS conditions, respectively (more NLOS locations were measured at lower frequencies).}
	\label{fig:140PL}
	\vspace{-0.5cm}
\end{figure}

Fig. \ref{fig:140PL} shows the scatter plot of synthesized omnidirectional measured path loss (without antenna gains) at 28, 38, 73, and 142 GHz in outdoor UMi environments ranging from 20-930 m, where the LOS and NLOS path loss data are denoted as circles and diamonds, respectively, and different frequency bands are marked with various colors \cite{Sun16b,Mac15c,Samimi15b,rappaport2015wideband,Mac19COMP,xing21icc}.

\begin{table*}[!ht]
	\centering
	\caption{Directional UMi CI and CIF path loss models and RMS delay spread in both LOS (boresight), NLOS Best, and NLOS Arbitrary pointing directions at 28 GHz (Manhattan, 31-187m), 38 GHz (Austin, 29-930m), 73 GHz (Brooklyn, 21-170m), and 142 GHz (Brooklyn, 24-117m) \cite{rappaport2015wideband, Mac19COMP,Mac15c,Sun16b,Samimi15b}. The weighted average frequency $f_0$ is 73 GHz for LOS scenarios and 62 GHz for NLOS scenarios. }~\label{tab:PLcomp}
	\begin{tabular}{?c|c?c|c|c|c?c|c|c|c?c|c|c|c?}
		\hline
		\multicolumn{2}{?c?}{\textbf{Urban Microcell}}                              & \multicolumn{4}{c?}{\textbf{Directional-LOS}} & \multicolumn{4}{c?}{$\textbf{Directional-NLOS}_{\textbf{Best}}$} & \multicolumn{4}{c?}{\textbf{Directional-NLOS}} \\ \hline
		\multicolumn{2}{?c?}{\textbf{Frequency [GHz]}}                       & \textbf{28 } & \textbf{38}  & \textbf{73  }& \textbf{142 }   & \textbf{28 } & \textbf{38 }& \textbf{73 }  & \textbf{142}  & \textbf{28 } & \textbf{38 }  & \textbf{73}   & \textbf{142} \\ \hline
		\multicolumn{2}{?c?}{\textbf{Directional Antenna HPBW}}                       & 11\textdegree & 8\textdegree   & 7/15\textdegree & 8\textdegree   & 11\textdegree & 8\textdegree  & 7/15\textdegree  & 8\textdegree  & 11\textdegree & 8\textdegree   & 7/15\textdegree   & 8\textdegree \\ \hline
		
		\multirow{1}{*}{\textbf{{Single-Band}}}      & $n$                        & 2.3 & 1.9  & 2.0       & 2.1   & 3.8 & 2.7 & 3.1    & 3.1   & 4.5 & 3.3  & 4.6   & 3.60       \\ \cline{2-14} 
		\multirow{1}{*}{\textbf{{PL CI }}} & $\sigma^{CI}$ {[}dB{]} & 4.3  & 3.5 & 1.9     & 2.8  & 9.3 & 7.9 & 10.5     & 8.3   & 10.0 & 10.3   & 10.5   & 9.1      \\ \hline \hline
		
		\multirow{1}{*}{\textbf{Multi-Band}}      & $n$    &\multicolumn{4}{c?}{2.07} & \multicolumn{4}{c?}{3.21}  & \multicolumn{4}{c?}{3.96} \\ \cline{2-14} 
		\multirow{1}{*}{\textbf{PL CI}}   & $\sigma^{CI}$ {[}dB{]} & \multicolumn{4}{c?}{3.6}  & \multicolumn{4}{c?}{9.8}& \multicolumn{4}{c?}{11.5} \\ \hline \hline
		
		\multirow{1}{*}{\textbf{Multi-Band}}      & $n, b$    &\multicolumn{4}{c?}{$n= 2.07, b= -0.10$} & \multicolumn{4}{c?}{$n= 3.21, b= -0.03$}  & \multicolumn{4}{c?}{$n= 3.96, b= -0.05$} \\ \cline{2-14} 
		\multirow{1}{*}{\textbf{PL CIF}}   & $\sigma^{CIF}$ {[}dB{]} & \multicolumn{4}{c?}{3.5}  & \multicolumn{4}{c?}{9.6}& \multicolumn{4}{c?}{11.5} \\ \hline \hline

		\multirow{4}{*}{\textbf{{RMS DS} \cite{rappaport2015wideband}}} 
		& $\min_{DS}$   [ns]            & 0.8  & N/A   & 0.7         & 0.7     & 1.0   &  N/A         & 0.6     &    0.6   & 0.5   & 1.0 & 0.5   & 0.6 \\
		\cline{2-14} 
		& $\max_{DS}$   [ns]            & 2.6 & N/A   & 0.7 & 13.9  & 165.1 & N/A  & 77.0 & 32.7  & 420.0 & 180.0  & 290.1 & 53.0 \\ \cline{2-14} 
		& $\mu_{DS}$ [ns]               & 0.9 & N/A          & 0.7 & 1.7  & 17.9 & N/A & 10.3& 4.5  & 25.7 & 11.4  & 23.4 & 9.2          \\ \cline{2-14} 
	& $\sigma_{DS}$   [ns]            & 1.0 & N/A  & 0.1 & 3.4  & 13.0& N/A  &18.7 & 9.7  & 25.0 & N/A  & 31.6 & 17.4 \\ 
		\hline

	\end{tabular}
\vspace{-0.5cm}
\end{table*}


The UMi LOS omnidirectional PLEs at 28, 38, 73, and 142 GHz are $n=$ 2.1, 1.9, 1.9, and 1.9, respectively, when referenced to a 1 m free space propagation distance, with $\sigma =$ 3.6 dB, 4.4 dB, 1.7 dB, and 2.7 dB, which are all slightly lower than the LOS directional PLEs at those frequencies. The LOS omnidirectional channels are seen to be virtually identical over all the frequencies and offer 1-2 dB less average path loss per decade of distance than the LOS directional channels (with antenna gains removed). 



The UMi NLOS omnidirectional PLEs are $n=$ 3.4, 2.7, 2.8, and 2.9 at 28, 38, 73, and 142 GHz, respectively, when referenced to the first-meter free space path loss, which are much smaller than the arbitrary directional NLOS PLEs but close to the $\text{NLOS}_{\text{Best}}$ directional PLEs, as shown in Tables \ref{tab:PLcomp} and \ref{tab:Omnicomp}, indicating that the $NLOS_{Best}$ pointing direction offers the single dominant propagation path among all pointing directions in NLOS scenarios. \textcolor{black}{The higher PLEs at 28 GHz are likely due to the dense and busy measurement environment in Manhattan which provided more obstructions, and there were fewer TX pointing angles measured at 28 GHz \cite{rappaport2013millimeter} which overestimated the omnidirectional path loss.}

Notably, the arbitrary directional NLOS channels are much lossier than the $NLOS_{Best}$ channel, indicating the need for narrow beam directional antennas in all mmWave and sub-THz systems. Thus, futuristic mmWave and THz communication systems will need fast and accurate beamforming algorithms to find, capture, and combine the most dominant multipath to maintain and extend the outdoor NLOS communication range at frequencies above 100 GHz \cite{Sun14a}. Both the LOS and NLOS omnidirectional PLEs at 38, 73, and 142 GHz are remarkably close to each other and are slightly lower than the PLEs at 28 GHz, showing that the path loss models from 28 GHz to 142 GHz are very similar regarding the PLEs when referenced to the first-meter propagation.

\begin{table*}[]
	\centering
	\caption{\textcolor{black}{Omnidirectional UMi path loss CI and CIF models, RMS angle of arrival spread (ASA) and RMS angle of departure spread (ASD)  at 28 GHz (Manhattan, 31-187 m), 38 GHz (Austin, 29-930 m), 73 GHz (Brooklyn. 21-170 m), and 142 GHz (Brooklyn, 24-117m) compared to the 3GPP UMi standard (CI model) \cite{Mac15c,rappaport2015wideband,Mac19COMP,Sun16b,Samimi15b,3GPP2019}. The weighted average frequency $f_0$ of CIF path loss models is 73 GHz for LOS  and 62 GHz for NLOS scenarios. \textcolor{black}{A 30 dB down threshold from the peak MPC power at each RX location was used to detect MPCs.} }}~\label{tab:Omnicomp}
	\begin{tabular}{?c|c?c|c|c|c|c?c|c|c|c|c?}
		\hline
		\multicolumn{2}{?c?}{\textbf{Urban Microcell}}                       & \multicolumn{5}{c?}{\textbf{Omnidirectional-LOS}} & \multicolumn{5}{c?}{\textbf{Omnidirectional-NLOS}} \\ \hline
		
		\multicolumn{2}{?c?}{\textbf{Omnidirectional NYU \& 3GPP}}                       & \multicolumn{4}{c|}{\textbf{NYU WIRELESS}} &\multicolumn{1}{c?}{\textbf{3GPP UMi CI}} &\multicolumn{4}{c|}{\textbf{NYU WIRELESS}} &\multicolumn{1}{c?}{\textbf{3GPP UMi CI}}\\ \hline
		\multicolumn{2}{?c?}{\textbf{Frequency [GHz]}}                    & \textbf{28}  & \textbf{38}   &\textbf{ 73}      & \textbf{142}  & \textbf{0.5-100}    & \textbf{28} &\textbf{38}     & \textbf{73}      & \textbf{142}  &\textbf{0.5-100}  \\ \hline
		
		\multirow{1}{*}{\textbf{Single-Band}}      & $n$    & 2.1 & 1.9 & 1.9   & 1.9  & N/A   & 3.4 & 2.7  & 2.8    & 2.9  & N/A   \\ \cline{2-12} 
		\multirow{1}{*}{\textbf{PL CI}}  & $\sigma^{CI}$ {[}dB{]} & 3.6  & 4.4  & 1.7   & 2.7  & N/A& 9.7 & 10.1  & 8.7   & 8.2  & N/A \\ \hline \hline
		
		\multirow{1}{*}{\textbf{Multi-Band}}      & $n$    &\multicolumn{4}{c|}{$1.91$} & \multicolumn{1}{c?}{2.1}  & \multicolumn{4}{c|}{2.96}& \multicolumn{1}{c?}{3.2} \\ \cline{2-12} 
		\multirow{1}{*}{\textbf{PL CI }}   & $\sigma^{CI}$ {[}dB{]} & \multicolumn{4}{c|}{3.72}  & \multicolumn{1}{c?}{4.0}& \multicolumn{4}{c|}{10.93}&\multicolumn{1}{c?}{8.2}    \\ \hline \hline
		
		\multirow{1}{*}{\textbf{Multi-Band}}      & $n, b$    &\multicolumn{4}{c|}{$n= 1.91, b = -0.03$}  & N/A  & \multicolumn{4}{c|}{\textbf{$n= 2.96, b = -0.05$}} & N/A  \\ \cline{2-12} 
		\multirow{1}{*}{\textbf{PL CIF }}   & $\sigma^{CIF}$ {[}dB{]} & \multicolumn{4}{c|}{3.69 }   & N/A& \multicolumn{4}{c|}{10.71}   & N/A    \\ \hline \hline
		
		\multirow{4}{*}{\textbf{ RMS AOA spread }} & $\min_{ASA}$ [ns] & 0\textdegree     & N/A    & 8.8\textdegree &3.2\textdegree & N/A    & 2.6\textdegree     & N/A  & 15.3\textdegree     & 3.4\textdegree& N/A  \\ \cline{2-12} 
			& $\max_{ASA}$  [ns]            & 58.4\textdegree      & N/A   &36.3\textdegree     &15.3\textdegree  & N/A  &62.2\textdegree     & N/A    & 65.6\textdegree & 59.2\textdegree & N/A \\ \cline{2-12} 
			& $\mu_{ASA}$ [ns]                 & 30.8\textdegree     & N/A     & 19.3\textdegree  & 10.1\textdegree  & N/A  & 32.5\textdegree  & N/A  & 33.5\textdegree  &32.5\textdegree &N/A \\ \cline{2-12} 
			& $\sigma_{ASA}$  [ns]            & 26.2 \textdegree   & N/A   &8.9\textdegree & 3.1\textdegree  & N/A   & 23.8\textdegree    & N/A   & 12.3\textdegree & 18.2 \textdegree &N/A    \\ \hline \hline
			
			\multirow{4}{*}{\textbf{ RMS AOD spread }} & $\min_{ASD}$ [ns] & 0\textdegree     & N/A    & 3.2\textdegree &0.6\textdegree & N/A    &4.0\textdegree     & N/A  & 7.0\textdegree     & 0\textdegree& N/A  \\ \cline{2-12} 
			& $\max_{ASD}$  [ns]            & 42.9\textdegree      & N/A   &10.8\textdegree     &21.7\textdegree  & N/A  &40.4\textdegree     & N/A    & 33.7\textdegree & 18.0\textdegree & N/A \\ \cline{2-12} 
			& $\mu_{ASD}$ [ns]                 & 12.5\textdegree     & N/A     & 5.3\textdegree  & 6.0\textdegree  & N/A  & 22.4\textdegree  & N/A  & 15.8\textdegree  &6.3\textdegree &N/A \\ \cline{2-12} 
			& $\sigma_{ASD}$  [ns]            & 16.0 \textdegree   & N/A   &2.4\textdegree & 5.3\textdegree  & N/A   & 12.0\textdegree    & N/A   & 8.4\textdegree & 6.5 \textdegree &N/A    \\ \hline \hline
			
			\multirow{1}{*}{\textbf{Number of AOA}}      & $n_{AOA}$    & 3.6& N/A & 2.8   & 1.9 & N/A   & 4.7 &  N/A  & 4.3    & 4.1  & N/A   \\ \cline{2-12} 
			\multirow{1}{*}{\textbf{Directions}}  & $\sigma_{AOA}$  & 3.4  & N/A  & 3.2   & 1.1 & N/A& 3.0 &  N/A & 2.8   & 2.6  & N/A \\ \hline  \hline
			
			\multirow{1}{*}{\textbf{Number of  AOD}}      & $n_{AOD}$    & 2.1 &  N/A & 1.6   & 1.3 & N/A   & 3.3 &  N/A  & 2.2   & 1.6  & N/A   \\ \cline{2-12} 
			\multirow{1}{*}{\textbf{Directions}}  & $\sigma_{AOD}$  & 2.6  &  N/A & 1.0  & 1.3  & N/A& 2.5 &  N/A  & 1.9  & 2.1  & N/A \\ \hline 
		
		
	\end{tabular}
	\vspace{-0.5cm}
\end{table*}

\subsection{UMi CI and CIF Models in Multi-Frequency Bands}

The multi-frequency CI model with a frequency-weighted PLE (CIF) \cite{Mac15b,Sun16b} was proposed as a viable multi-band path loss model valid across various frequency bands \cite{Xing21b,Rap17a,Sun16b}:
\begin{equation}
\label{equ:CIF}
\small
\begin{split}
PL^{CIF}(f_c, &d_{\text{3D}}) = \text{FSPL}(f_c, d_0)+\\
  &10n\left( 1+ b \left( \dfrac{f-f_0}{f_0}\right) \right) \log_{10}\left( \dfrac{d}{d_0} \right) + \chi^{CIF}_{\sigma},\\
 & \text{where~} f_0=\sum_{k=1}^{K}f_kN_k/\sum_{k=1}^{K}N_k.
\end{split}
\end{equation}
In addition to the PLE $n$, the two-parameter CIF path loss model \eqref{equ:CIF} uses another parameter $b$ to present the frequency dependency of path loss, and a large absolute value of $b$ (e.g., $b=$ 0.5) means the PLE is highly dependent on frequencies, vice versa.

Table \ref{tab:PLcomp} and \ref{tab:Omnicomp} summarize directional and omnidirectional multi-band CI and CIF path loss models for both LOS and NLOS scenarios, and both models provide nearly identical results in terms of the path loss and shadow fading over four frequency bands, as also shown in Fig. \ref{fig:140PL}. The 3GPP UMi omnidirectional path loss models \cite{3GPP2019} which are proposed for frequencies from 0.5-100 GHz are also presented in Table \ref{tab:Omnicomp} for comparisons. In general, the 3GPP UMi CI path loss models are close to the multi-frequency path loss models presented in this letter in terms of PLEs and shadow fading standard deviations in both LOS and NLOS scenarios. The PLEs of the 3GPP UMi path loss models are close to the PLEs at 28 GHz and overestimate the path loss at higher frequencies, which may be due to the fact that 3GPP models are developed from a large number of measurements around 28 GHz but few measurements at higher frequencies up to 100 GHz \cite{Rap17a}.

\section{RMS Delay Spreads at 28, 38, 73, and 142 GHz} \label{sec:DS}
The minimum, maximum, and mean of RMS delay spreads from the aforementioned outdoor UMi measurements at 28, 38, 73, and 142 GHz are presented in Table \ref{tab:PLcomp} for directional outdoor UMi channels \cite{rappaport2015wideband,rappaport2013millimeter,rappaport2013broadband}. A 5 dB SNR threshold was used in this letter for all four bands to detect and keep the MPCs in each power delay profile \cite{rappaport2015wideband}.

Table \ref{tab:PLcomp} shows that the UMi directional LOS RMS DS is negligible with a mean of  1-2 ns from 28 to 142 GHz, indicating the fact that the LOS multipath component is the only dominant path (no other MPCs or powers of other MPCs are negligible) when narrow-beam directional antennas are pointing in the boresight direction in LOS scenarios in mmWave and sub-THz bands. 

For UMi directional $\text{NLOS}_{\text{Best}}$ scenarios, the mean RMS DS ($\mu_{DS}$) is 18 ns, 10 ns, and 5 ns at 28, 73, and 142 GHz, respectively, which implies more than one multipath components can be received in NLOS scenarios with directional antennas. The maximum RMS DS ($\max_{DS}$) decreases as frequency increases and is 165 ns at 28 GHz, 77 ns at 73 GHz, and 32 ns at 142 GHz, with only 10\% of the RMS DS larger than 88 ns, 37 ns, and 15 ns in 28, 73, and 142 GHz bands, respectively. The mean and maximum RMS delay spread decreases as frequency increases, which is likely due to there are higher losses introduced by obstructions and foliage, and fewer reflections captured by the directional antennas at higher frequencies. 



When the RX antenna was not pointing to the $\text{NLOS}_{\text{Best}}$ direction in NLOS scenarios or to the LOS boresight direction in LOS scenarios (arbitrarily pointing), the maximum delay spread was observed to be 420 ns, 180 ns, 290 ns, and 53 ns at 28, 38, 73, and 142 GHz, respectively, with only 10\%, 6\%, 6\%, and 1\% of the RMS DS larger than 50 ns at 28, 73, and 142 GHz bands \cite{Ju20a,Mac15b}, respectively. The mean RMS delay spread is 26 ns at 28 GHz, 11 ns at 38 GHz, 23 ns at 73 GHz, and 9 ns at 142 GHz, which are all larger than the RMS delay spreads in $\text{NLOS}_{\text{Best}}$ and LOS boresight scenarios across all four frequencies bands, since for UMi NLOS scenarios there is likely more than one dominant multipath. The lower RMS DS at 38 GHz is due to there are less dense buildings and obstructions in Austin as compared to Manhattan and Brooklyn. The dropping RMS DS at higher frequencies indicates fewer multipath components at higher frequencies, which is likely because of the higher loss introduced by obstructions and foliage. 



\begin{figure}    
	\centering
	\includegraphics[width=0.45\textwidth]{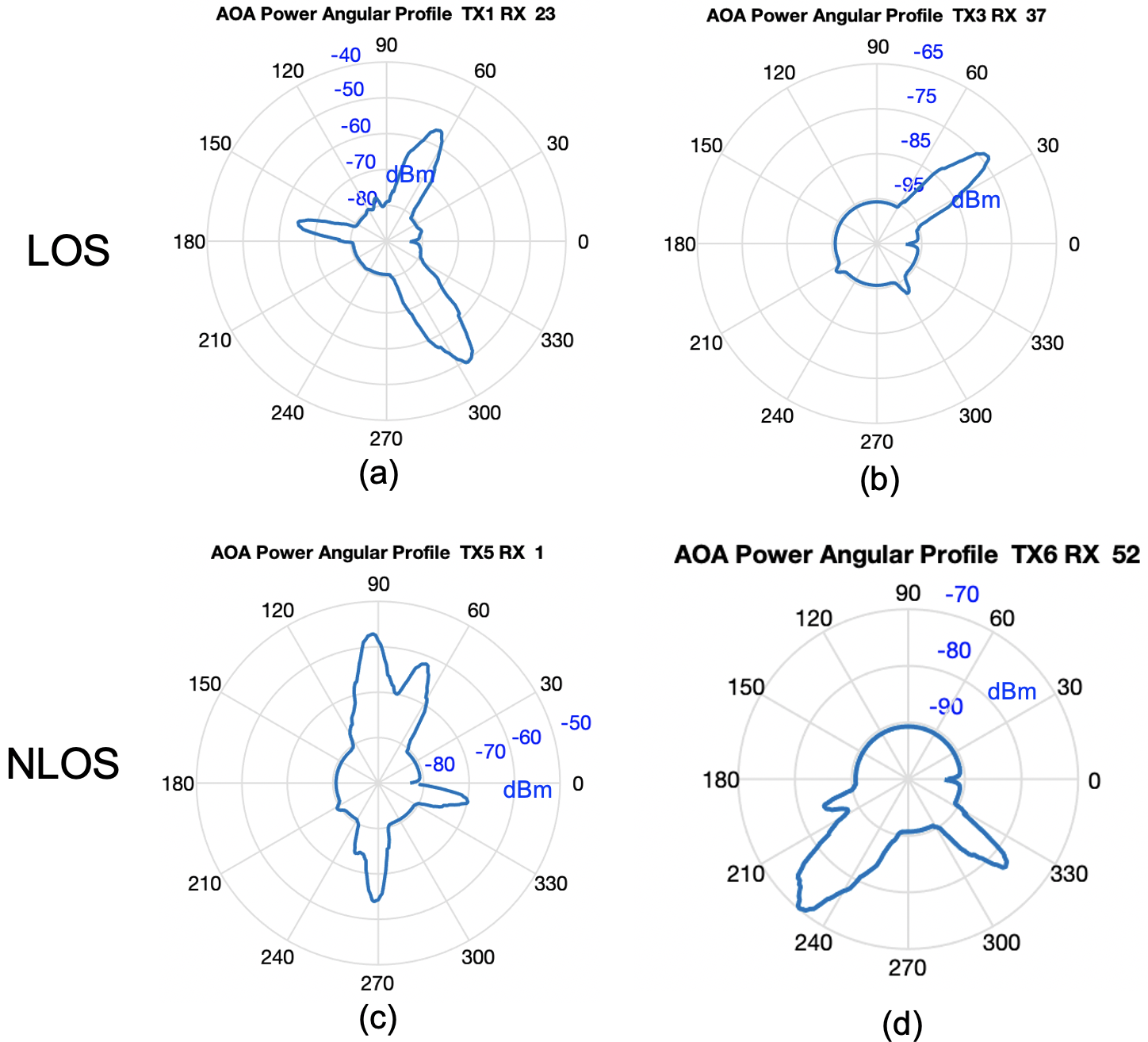}
	\caption{\textcolor{black}{142 GHz UMi Sample Power Angular Profiles in both LOS and NLOS scenarios.}}
	\label{fig:140AOA}
	\vspace{-0.5cm}
\end{figure}

\section{Angular Statistics at 28, 38, 73, and 142 GHz} \label{sec:AS}
\textcolor{black}{As wireless channels become more sparse and antenna beamwidth is narrower at higher frequencies, rapid beamforming and beam tracking algorithms are required in MIMO systems, which require accurate channel angular information. Channel angular statistics including the RMS angle of arrival (AOA) spread, RMS angle of departure (AOD) spread, and the average number of AOA and AOD directions extracted from the aforementioned outdoor UMi measurements at 28, 38, 73, and 142 GHz are summarized in Table \ref{tab:Omnicomp} \cite{rappaport2015wideband,rappaport2013millimeter,rappaport2013broadband}. A 30 dB down from the peak power threshold was used in this letter at 28, 73, and 142 GHz to detect and keep the MPCs in each power angular profile. }

\textcolor{black}{Fig. \ref{fig:140AOA} presents four sample AOA power angular profiles in both LOS and NLOS scenarios at 142 GHz. There are three and one distinct AOA directions observed in LOS scenarios as shown in Fig.  \ref{fig:140AOA} (a) and (b), respectively, and there are four and two separated AOA directions observed in NLOS scenarios as shown in Fig.  \ref{fig:140AOA} (c) and (d). Counting the number of AOA and AOD directions for each location pair at different frequencies, the average number AOA and AOD directions are presented in Table \ref{tab:Omnicomp}.}

\textcolor{black}{The mean of RMS AOA and AOD spread tends to decrease with increasing frequencies in both LOS and NLOS scenarios, and at the same frequency, the mean of RMS AOA spread is generally larger than the mean of RMS AOD spread, indicating there are more AOA directions than AOD directions. In our measurements, relatively the same beamwidth antennas (7-10\textdegree) were used at different frequencies as shown in Table \ref{tab:sounder}, and any spatial paths within the antenna beamwidth could not be resolved (e.g., 8\textdegree). If a narrower beam antenna was used, there might be more AOA or AOD directions resolved, and if a wider beam antenna was used, some AOA or AOD directions might be lost (not resolvable). However, the angular statistics will not change with the transmission bandwidth unless the bandwidth is huge (e.g., 10 or 20 GHz) which increases the thermal noise of the channel, and some weak spatial directions will not be detected. }

The number of AOA and AOD directions decreases with the increasing frequency in both LOS and NLOS scenarios, which shows the wireless channels are more sparse at higher frequencies. \textcolor{black}{Threshold level greatly impacts statistics such as the number of AOA/AOD directions but has negligible impact on path loss and RMS time/angular spreads (stronger MPCs contribute more). If a power threshold of 20 dB down from the peak MPC is used, the mean number of AOA directions at 142 GHz decreases to 1.7 and 2.8 in LOS and NLOS scenarios \cite{Ju21Globecom}, respectively, since some of the AOA directions with weak powers are filtered out by thresholding.} The spatial statistics (e.g., RMS angular spread, number of distinct AOA and AOD directions) will be useful for MIMO channel rank (the number of beams that can be supported by the MIMO matrix) analysis.

\section{Conclusion}\label{conclusion}
Comparisons of outdoor wireless channels (LOS and NLOS) in UMi environments at 28, 38, 73, and 142 GHz are presented in this letter, based on four different measurement campaigns conducted from 2011-2020 in downtown Austin, Manhattan, and Brooklyn. The path loss components of both CI and CIF models in both LOS and NLOS scenarios are notably similar over four frequency bands from mmWave to sub-THz frequencies, when referenced to the free space path loss in the first meter (near field) \cite{Sun16b,Mac15b,rappaport2013millimeter}, implying the outdoor UMi channels at THz frequencies are not that different from today's mmWave channels with the exception of the first meter path loss. The RMS delay spread and angular spread in outdoor UMi environments decrease as frequency increases when using directional antennas from 28 to 142 GHz. The comparisons of channel characteristics and models in this letter will support outdoor multi-band wireless system designs above 100 GHz for 6G and beyond.

\bibliographystyle{IEEEtran}
\bibliography{Indoor140GHznew}

\end{document}